\documentclass{article}

\usepackage{arxiv}

\usepackage[utf8]{inputenc} 
\usepackage[T1]{fontenc}    
\usepackage{hyperref}       
\usepackage{url}            
\usepackage{booktabs}       
\usepackage{amsfonts}       
\usepackage{nicefrac}       
\usepackage{microtype}      
\usepackage{lipsum}		
\usepackage{graphicx}
\usepackage{natbib}
\usepackage{doi}

\usepackage{amsmath,amssymb,amsfonts}
\usepackage{algorithmic}
\usepackage{graphicx}
\usepackage{textcomp}
\usepackage{xcolor}
\usepackage{array}
\usepackage{tabularx}
\usepackage{caption}
\usepackage[utf8]{inputenc}
\usepackage[export]{adjustbox}
\usepackage{subfig}

\usepackage{url}

\newcommand{\quotes}[1]{``#1''}

\setcitestyle{square}

\title{\Large{\textbf{COVID-19's (mis)information ecosystem on Twitter}}\\\large{How partisanship boosts the spread of conspiracy narratives on German speaking Twitter}}

\author{ \href{https://orcid.org/0000-0002-8751-0518}{\includegraphics[scale=0.06]{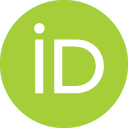}\hspace{1mm}Morteza Shahrezaye} \\
	Institute for Media and Communications Management\\
	University of St.Gallen\\
	\texttt{morteza.shahrezaye@unisg.ch} \\
	\And
	{\hspace{1mm}Miriam Meckel} \\
	Institute for Media and Communications Management\\
	University of St.Gallen\\
	\texttt{miriam.meckel@unisg.ch} \\
	\AND
	{\hspace{1mm}Léa Steinacker} \\
	University of St.Gallen\\
	\texttt{lea.steinacker@student.unisg.ch} \\
    \AND
	{\hspace{1mm}Viktor Suter} \\
	Institute for Media and Communications Management\\
	University of St.Gallen\\
	\texttt{viktor.suter@unisg.ch} \\
}

\date{}

\renewcommand{\headeright}{}
\renewcommand{\undertitle}{}
\renewcommand{\shorttitle}{}

\hypersetup{
pdftitle={COVID-19's (mis)information ecosystem on Twitter: How partisanship boosts the spread of conspiracy narratives on German speaking Twitter},
pdfauthor={Morteza Shahrezaye~ Miriam Meckel~ Léa Steinacker~ Viktor Suter},
pdfkeywords={social networks, misinformation, conspiracy theory, political polarization},
}

\begin{document}
\maketitle

\begin{abstract}
In late 2019, the gravest pandemic in a century began spreading across the world. A state of uncertainty related to what has become known as SARS-CoV-2 has since fueled conspiracy narratives on social media about the origin, transmission and medical treatment of and vaccination against the resulting disease, COVID-19. Using social media intelligence to monitor and understand the proliferation of conspiracy narratives is one way to analyze the distribution of misinformation on the pandemic. We analyzed more than 9.5M German language tweets about COVID-19. The results show that only about 0.6\% of all those tweets  deal with conspiracy theory narratives. We also found that the political orientation of users correlates with the volume of content users contribute to the dissemination of conspiracy narratives, implying that partisan communicators have a higher motivation to take part in conspiratorial discussions on Twitter. Finally, we showed that contrary to other studies, automated accounts do not significantly influence the spread of misinformation in the German speaking Twitter sphere. They only represent about 1.31\% of all conspiracy-related activities in our database. 
\end{abstract}

\keywords{social networks\and misinformation\and conspiracy theory\and political polarization}

\section{Introduction}
In November 2019, a febrile respiratory illness caused by SARS-CoV-2 infected people in the city of Wuhan, China. 
On January 30th 2020, the World Health Organization (WHO) declared the spread of the virus a worldwide pandemic\citep{bc.com/news/world-51839944}. Shortly after, the WHO reported multiple COVID-19-related knowledge gaps relating to its origin, transmission, vaccinations, clinical considerations, and concerns regarding the safety of healthcare workers\citep{who.int/who-documents-detail/}. The organization warned of an \quotes{infodemic}, defined by \quotes{an overabundance of information and the rapid spread of misleading or fabricated news, images, and videos}\citep{who.int/news-room/feature-stories/detail}. By August 2020, more than 22 million people worldwide had contracted the virus\citep{who.int/docs/default-source/coronaviruse}. The Organization for Economic Co-operation and Development (OECD) put forward estimates of negative GDP growth for all member countries in 2020 due to the crisis \citep{OECD7969896b-en}.  

COVID-19's indomitable dissemination around the globe combined with a lack of effective medical remedies \citep{guo2020origin,10.3389/fpubh.2020.00189} and its psychological and economic side effects \citep{OECD7969896b-en,ho2020mental,rajkumar2020covid,frankcovid} have left many people to uncertainty and fear of further developments. Previous research has shown that lack of certainty and control often results in the emergence and circulation of conspiracy theory narratives \citep{Whitson115}. Popper defined conspiracy mentality as the \quotes{mistaken theory that, whatever happens in society – especially happenings such as war, unemployment, poverty, shortages, which people as a rule dislike – is the result of direct design by some powerful individuals and groups}\citep{Popper2002}. This one-sided or even pathological method of reasoning regularly facilitates coping with uncertainty and fear by making the world more understandable and providing individuals with an illusion of control \citep{kruglanski2006groups}. 


There are two main conditions conducive to the emergence of conspiracy narratives: individuals' psychological traits and socio-political factors. Regarding psychological traits, numerous laboratory studies demonstrate the correlation between conspiracy beliefs and psychological features like negative attitude toward authorities \citep{doi:10.1002/per.1930}, self-esteem \citep{doi:10.1111/0162-895X.00160}, paranoia and threat \citep{doi:10.1080/13608746.2017.1359894}, powerlessness \citep{doi:10.1111/0162-895X.00160}, education, gender and age \citep{doi:10.1002/acp.3301}, level of agreeableness \citep{doi:10.1111/j.2044-8295.2010.02004.x}, and death-related anxiety \citep{NEWHEISER20111007}. Another part of reasoning sees conspiracy mentality as a generalized political attitude \citep{doi:10.1002/per.1930} and correlates conspiracy beliefs to socio-political factors like political orientation. Enders et al. showed that conspiracy beliefs can be a product of partisanship \citep{enders_smallpage_lupton_2020}. Several other studies show a quadratic correlation between partisanship and the belief in certain conspiracy theories  \citep{doi:10.1177/1948550614567356}. These insights imply that extremists on both sides of the political spectrum are more prone to believe in and to discuss conspiracy narratives.  

We define conspiracy narratives as part of the overall phenomenon of misinformation on the internet. We use misinformation as the broader concept of fake or inaccurate information that is not necessarily intentionally produced (distinguished from disinformation which is regularly based on the intention to mislead the recipients). Among all the conspiracy narratives, we are interested in those propagated in times of pandemic crises. The spread of health-related conspiracy theories is not a new phenomenon \citep{10.2307/43287281,bogart2010conspiracy,klofstad2019drives} but seems to be even accelerated in world connected via social media.   
                 
The COVID-19 pandemic's unknown features, its psychological and economical side effects, the ubiquitous availability of Online Social Networks (OSNs)\citep{who.intsss/docs/default-source/coronaviruse}, and high levels of political polarization in many countries \citep{doi:10.1177/1940161219892768,10.1111/jcc4.12166} make this pandemic a potential breeding ground for the spread of conspiracy narratives. From the outset of the crisis,  \quotes{misleading rumors and conspiracy theories about the origin circulated the globe paired with fear-mongering, racism, and the mass purchase of face masks[...]. The social media panic travelled faster than the COVID-19 spread} \citep{10.1093/jtm/taaa031}. Such conspiracy narratives can obstruct the efforts to properly inform the general public via medical and scientific findings \citep{10.1371/journal.pone.0147905}. Therefore, investigating the origins and circulation of conspiracy narratives as well as the potential political motives supporting their spread on OSNs is of vital public relevance.  With this objective, we analyzed more than 9.5M German language tweets about COVID-19 to answer the following research questions:

\textbf{Research Question 1}:
What volume of German speaking Twitter activities comprises COVID-19 conspiracy discussions and how much of this content is removed from Twitter?

\textbf{Research Question 2}:
Does the engagement with COVID-19 conspiracy narratives on Twitter correlate with political orientation of users?


\textbf{Research Question 3}:
To what degree do automated accounts contribute to the circulation of conspiracy narratives in the German speaking Twitter sphere? 

\section{Data}
We collected the data for this study during the early phase of the crisis, namely, between March 11th, the day on which the WHO declared the spread of the SARS-CoV-2 virus a pandemic\citep{bc.com/news/world-51839944} and May 31st, 2020. The data was downloaded using the Twitter's Streaming API by looking for the following keywords: \quotes{COVID}, \quotes{COVID-19}, \quotes{corona}, and \quotes{coronavirus}. Only Tweets posted by German speaking users or with German language were included. The final dataset comprises more than 9.5M tweets from which two categories of conspiracy narratives were selected: conspiracy narratives about the origin of the COVID-19 illness (Table \ref{tab:origin}) and those about its potential treatments (Table \ref{tab:treatments}). The conspiracy narratives about the origin of the COVID-19 illness were selected based on Shahsavari et al., who automatically detected the significant circulation of the underlying conspiracy theories on Twitter using machine learning methods \citep{shahsavari2020conspiracy}. The second group of conspiracy narratives were chosen based on the fact that they were in the center of attention in German media \citep{TagesschauHomepage} and thus a considerable number of tweets discussed them \citep{NetzpolitikHomepage}.

\begin{table}[!ht]
\caption{Conspiracy narratives about the origin of COVID-19}\label{tab:origin}
\begin{tabularx}{\linewidth}{|>{\hsize=0.23\hsize}X|>{\hsize=0.8\hsize}X|}
\hline
case             & description\\\hline
5G               & conspiracy narrative suggesting that the 5G network activates the virus\\\hline
Bill Gates       & conspiracy narrative suggesting that Bill Gates aims to use COVID-19 to initiate a global surveillance regime\\\hline
Wuhan laboratory & conspiracy theory narrative suggesting that the virus originates from a laboratory in Wuhan, China\\\hline
\end{tabularx}
\end{table}

\begin{table}[!ht]
\caption{Conspiracy narratives about potential treatments of COVID-19}\label{tab:treatments}
\begin{tabularx}{\linewidth}{|>{\hsize=0.23\hsize}X|>{\hsize=0.8\hsize}X|}
\hline
case        & description                                                                                      \\\hline
Ibuprofen   & conspiracy narrative suggesting that Ibuprofen reduces COVID-19 symptoms                        \\\hline
Homoeopathy & conspiracy narrative suggesting that homeopathy medicines reduce COVID-19 symptoms              \\\hline
Malaria     & conspiracy narrative suggesting that a malaria drug is an antiviral against SARS-CoV-2 virus \\\hline
\end{tabularx}
\end{table}

\begin{figure}[htbp]
\centerline{\includegraphics[width=\linewidth]{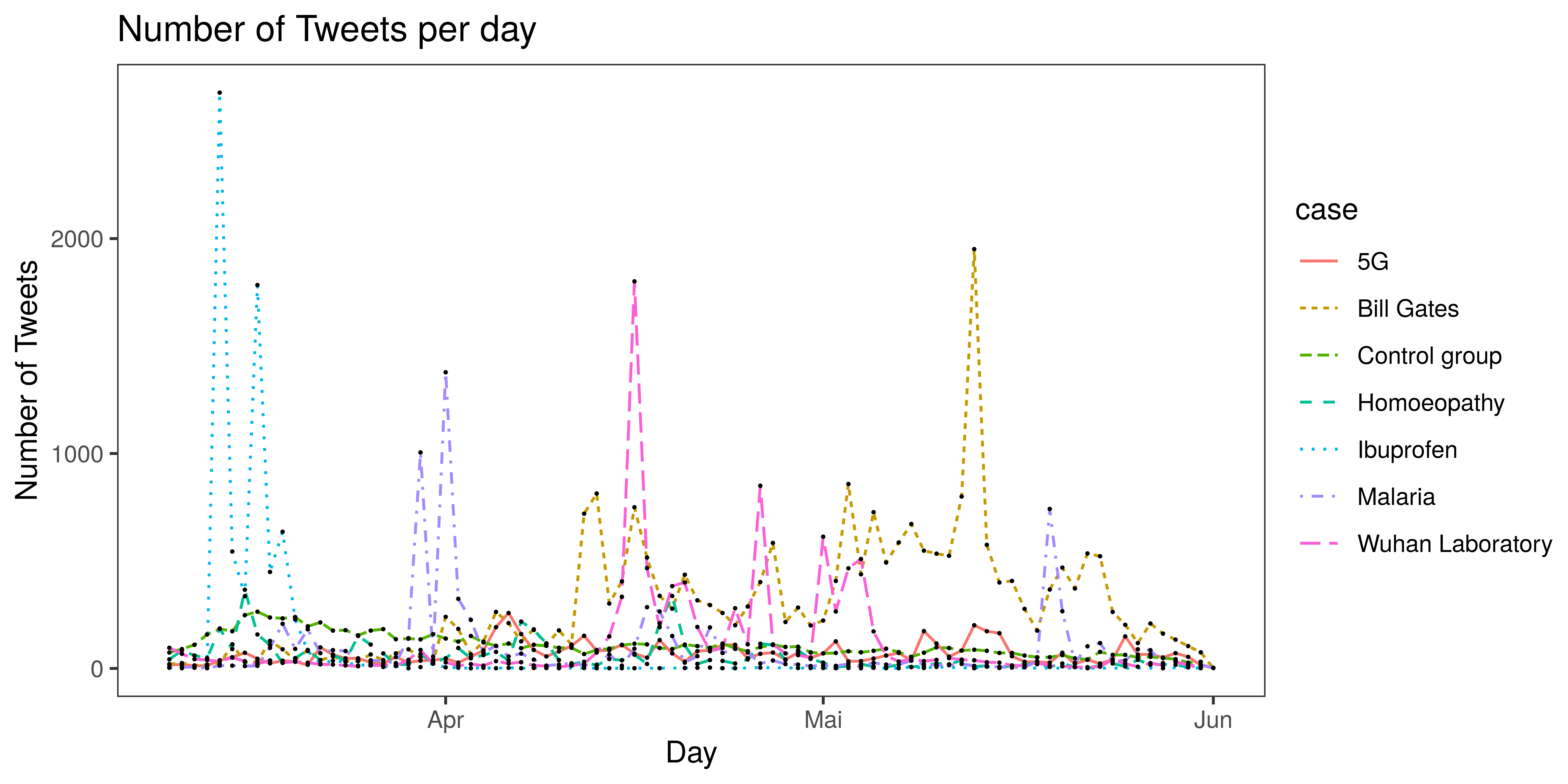}}
\caption{Daily number of tweets for each conspiracy narrative}\label{fig:timeline}
\label{fig}
\end{figure}

\noindent
Table \ref{tab:stats} indicates the number of tweets  belonging to each conspiracy narrative and the keywords that are used to filter them out\footnote{We used \quotes{homöopath} in order to match both German words \quotes{homöopathie} and \quotes{homöopathisch}}. There were 68,466 tweets in total discussing the underlying conspiracy narratives. Figure \ref{fig:timeline} shows the timeline of the tweets.

\begin{table}[!ht]
\caption{Number of tweets for each conspiracy narrative}\label{tab:stats}
\begin{tabularx}{\linewidth}{|X|X|X|}
\hline
case             &  number of tweets & keywords   \\\hline
5G               &  5,762            & 5G, \#5g                    \\\hline
Bill Gates       &  24,653           & Bill Gates, \#billgates     \\\hline
Wuhan laboratory &  9,366            & \#wuhanlab Wuhan Lab        \\\hline
Ibuprofen        &  7,016            & Ibuprofen, \#ibuprofen      \\\hline
Homeopathy      &  4,714            & Homöopath, \#Homöopath      \\\hline
Malaria          &  7,955            & Malaria, \#malaria          \\\hline
Control group    &  9,000            & \_\_                        \\\hline
\end{tabularx}
\end{table}

In addition to the six conspiracy narratives, 9000 tweets were randomly extracted from the dataset and served as a control group.

To answer research question 2, a list was extracted from official party websites; this list contains members of parliament (MPs) who are active on Twitter and belong to one of the six political parties in Germany's federal legislature. Each party runs several official Twitter pages that were added to the list of Twitter pages of each political party; for example, the official Twitter page of the Social Democratic Party (SPD) in the federal state of Bavaria, called \quotes{BayernSPD}, was added to the SPD list. For each twitter account in the extracted list a maximum of 4000 tweet handles were downloaded from the Twitter API. Table \ref{tab:partystats} shows the relevant statistics on the political tweets. 

\begin{table}[!ht]
\caption{Number of political tweets extracted from politicians' twitter pages}\label{tab:partystats}
\begin{tabularx}{\linewidth}{|X|X|X|X|}
\hline
political party       & number of MPs on Twitter & number of extra official Twitter pages & total number of tweet handles \\\hline
AfD                   & 27                           & 14                                     & 68,789                    \\\hline
CDU/CSU               & 131                          & 18                                     & 220,768                   \\\hline
FDP                   & 56                           & 7                                      & 96,046                    \\\hline
Bündnis 90/Die Grünen & 56                           & 11                                     & 169,864                   \\\hline
Linke                 & 50                           & 12                                     & 155,794                   \\\hline
SPD                   & 110                          & 17                                     & 221,029                   \\\hline
\end{tabularx}
\end{table}

In the next step, for each of the 68,466 users spreading conspiracy narratives (Table \ref{tab:stats}) the lists of their tweet handles were downloaded (Table \ref{tab:userstats}). Finally, for each of them we counted the number of times they retweeted one of the political tweets in table \ref{tab:partystats}. Based on Boyd et al. retweets are mainly a form of endorsement \citep{boyd2010tweet}. Therefore, we assume if a user collects a discernible number of retweets from members of a certain political party, this user will most likely share the corresponding political orientation. This method of inference about the political orientation of users has been applied in similar studies \citep{garimella2017mary}.

\begin{table}[!ht]
\caption{Number of tweets extracted from users spreading conspiracy narrative tweets}\label{tab:userstats}
\begin{tabularx}{\linewidth}{|X|X|X|}
\hline
case             & number of tweets & number of downloaded tweets from the contributing users \\\hline
5G               & 5762                     & 10,967,158                                              \\\hline
Bill Gates       & 24653                    & 35,144,536                                              \\\hline
Wuhan laboratory & 9366                     & 14,332,403                                              \\\hline
Ibuprofen        & 7016                     & 14,855,267                                              \\\hline
Homoeopathy      & 4714                     & 7,746,555                                               \\\hline
Malaria          & 7955                     & 16,258,164                                              \\\hline
Control group    & 9000                     & 12,217,082                                              \\\hline
\end{tabularx}
\end{table}

\section{Results}

There are multiple studies showing that exposure to misinformation can lead to persistent negative effects on citizens. The respondents in a study adjusted their judgment proportional to their cognitive ability after they realized that their initial evaluation was based on inaccurate information. In other words, respondents with lower levels of cognitive ability tend to keep biased judgments even after exposure to the truth \citep{DEKEERSMAECKER2017107}. In another study, Tangherlini
et al. found that conspiracy narratives stabilize based on the alignment of various narratives, domains, people, and places such that the removal of one or some of these entities would cause the conspiracy narrative to quickly fall apart \citep{10.1371/journal.pone.0233879}. Imhoff and Lamberty have shown that believing COVID-19 to be a hoax negatively correlated with compliance with self-reported, infection-reducing, containment-related behavior \citep{imhoff2020bioweapon}. 

On that account, to assess  a democratic information ecosystem that is balanced rather towards reliable information than misinformation we need to monitor and estimate if COVID-19 conspiracy theory narratives circulate significantly on Twitter. Based on a survey in mid-March 2020, about 48\% of respondents stated that they have seen some pieces of likely misinformation about COVID-19\citep{TagesschauHomepages}.  Shahsavari
et al. used automated machine learning methods to automatically detect COVID-19 conspiracy narratives on Reddit, 4Chan, and news data \citep{shahsavari2020conspiracy}. Multiple other studies found evidence of COVID-19 misinformation spread on different OSNs \citep{boberg2020pandemic,ahmed2020covid,serrano2020nlp}. 

To address the public concerns many of the service providers claimed that they will remove or tag this sort of content on their platforms. On March 16th 2020, Facebook, Microsoft, Google, Twitter and Reddit said they are teaming up to combat COVID-19 misinformation on their platforms\citep{Bloombergs}. On April 22nd, Twitter stated that they have removed over 2230 tweets containing misleading and potentially harmful COVID-19-related content\citep{BloombergTwitters}. On June 7th 2020, we examined how many of the German conspiracy-related tweets still exist on Twitter in order to understand if conspiracy-related tweets tend to exist on Twitter for a longer period of time compared to non conspiracy-related tweets. Table \ref{tab:contentmoderation} shows the results.

\begin{table}[!ht]
\caption{Share of conspiracy narratives among all COVID-19 tweets}\label{tab:contentmoderation}
\begin{tabularx}{\linewidth}{|X|X|X|X|}
\hline
case             & number of tweets  & share (among all 9.5M COVID-19 tweets) & share deleted on 7th June 2020 \\\hline
5G               & 5762                     & 0.06\%                                 & 6\%                            \\\hline
Bill Gates       & 24653                    & 0.25\%                                 & 7\%                            \\\hline
Wuhan laboratory & 9366                     & 0.098\%                                & 9\%                            \\\hline
Ibuprofen        & 7016                     & 0.073\%                                & 14\%                           \\\hline
Homoeopathy      & 4714                     & 0.049\%                                & 3\%                            \\\hline
Malaria          & 7955                     & 0.083\%                                & 5\%                            \\\hline
Control group    & 9000                     & 0.094\%                                & 6\%                            \\\hline
\end{tabularx}
\end{table}

\subsection{Research Question 1}
Based on Table \ref{tab:contentmoderation}, only about 0.61\% of all COVID-19 German tweets are about one of the conspiracy narratives under consideration. These German tweets are posted by more than 36,000 unique Twitter users. While 0.61\% is small in magnitude, it still comprises a relevant number of citizens. It is important to note though that this finding does not imply that only about 36,000 Twitter users believe in conspiracy theories. While our data shows the spread of conspiracy narratives, they do not reveal a user's stance towards the respective content.
In terms of content moderation by Twitter, on average 7.3\% of conspiracy narrative tweets are deleted after a certain period of time which is significantly higher than 6\% of tweets in the control group. We speculate that more of the conspiracy-related tweets are deleted because of Twitter's content moderation efforts that have been enforced due to recent public debates about misinformation on OSNs. 
 
\subsection{Research Question 2}
There is a long list of laboratory studies that show a correlation between conspiracy mentality and extreme political orientation \citep{enders_smallpage_lupton_2020,doi:10.1177/1948550614567356}. In this study we answer the slightly different question if the partisanship of Twitter users correlates with their contribution to conspiracy theory narrative discussions. Table \ref{tab:orientation} shows the distribution of the political orientations of users who discuss each of the underlying conspiracy narratives. 

\begin{table}[!ht]
\caption{Political orientation of users discussing conspiracy narratives}\label{tab:orientation}
\begin{tabularx}{\linewidth}{|X|X|X|X|X|X|X|X|}
\hline
case             & AfD  & CDU/ CSU & FDP  & Bündnis 90/Die Grünen & Linke & SPD  & Unknown \\\hline
5G               & 11\% & 3\%     & 3\%  & 8\%                   & 10\%  & 14\% & 51\%    \\\hline
Bill Gates       & 16\% & 3\%     & 3\%  & 8\%                   & 12\%  & 14\% & 44\%    \\\hline
Wuhan laboratory & 27\% & 3\%     & 15\% & 7\%                   & 5\%   & 8\%  & 35\%    \\\hline
Ibuprofen        & 9\%  & 3\%     & 3\%  & 8\%                   & 9\%   & 16\% & 52\%    \\\hline
Homoeopathy      & 5\%  & 4\%     & 6\%  & 13\%                  & 15\%  & 24\% & 33\%    \\\hline
Malaria          & 10\% & 2\%     & 2\%  & 5\%                   & 6\%   & 11\% & 64\%    \\\hline
Control group    & 10\% & 2\%     & 2\%  & 4\%                   & 6\%   & 10\% & 66\%    \\\hline
\end{tabularx}
\end{table}


\noindent
Table \ref{tab:orientation} demonstrates that users who are likely to be supporters of AfD and SPD most actively discuss and spread  COVID-related conspiracy narratives on Twitter. To check if contributions to conspiracy narratives are correlated with the political orientation of users, we ran a saturated Poisson log-linear model on the contingency Table \ref{tab:orientation}. The model defines the counts as independent observations of a Poisson random variable and includes the linear combination and the interaction between conspiracy narratives and the political orientation of users \citep{agresti2003categorical}.

\begin{equation}
log(\mu_{ij})=\lambda+\lambda^N_i+\lambda^P_j+\lambda^{NP}_{ij}
\end{equation}

\noindent
where $\mu_{ij}=E(n_{ij})$ represents the expected counts, $\lambda$s are parameters to be estimated and $N$ and $P$ stand for \textit{Narrative} and \textit{Political Orientation}. $\lambda^{NP}_{ij}$s corresponds to the interaction and association between conspiracy narratives and also reflects the departure from independence \citep{agresti2003categorical}. 
Since we suspect that beliefs in certain mutually contradictory conspiracy theories can be positively correlated \citep{wood2012dead}, we aggregated the six conspiracy theory cases to two based on to which category they belong and formed Table \ref{tab:orientation2} to remove any possible correlation.

\begin{table}[!ht]
\caption{Political orientation of users who discuss two conspiracy narratives (absolute counts)}\label{tab:orientation2}
\begin{tabularx}{\linewidth}{|X|X|X|X|X|X|X|X|}
\hline
case             & AfD  & CDU/ CSU & FDP  & Bündnis 90/Die Grünen & Linke & SPD  & Unknown \\\hline
Origins of COVID-19      & 4133  & 694     & 1396  & 1873                  & 2449  & 3028 & 10296   \\\hline
Possible treatments of COVID-19          & 1263 & 432     & 497  & 1149                   & 1347   & 2330 & 7841    \\\hline
\end{tabularx}
\end{table}

\noindent
Table \ref{tab:glm} shows the ANOVA analysis of the underlying saturated Poisson log-linear model applied on Table \ref{tab:orientation2}. The last line of resulting p-values in Table \ref{tab:glm} shows that in interaction parameter, $\mu_{ij}=E(n_{ij})$, is statistically significant. Therefore, we can reject the hypothesis that the contribution to conspiracy narratives is independent of the political orientation of users. The fact that there is evidence of a correlation between the contribution to conspiracy narratives and the political orientation of users, however, does not imply any causality.

\begin{table}[!ht]
\caption{ANOVA of Poisson log-linear model on the contingency Table \ref{tab:orientation}}\label{tab:glm}
\centering
\begin{tabular}{lrrrrr}
  \hline
 & Df & Deviance & Resid. Df & Resid. Dev & Pr($>$Chi) \\ 
  \hline
NULL &  &  & 13 & 31332.82 &  \\ 
  narrative & 1 & 2115.49 & 12 & 29217.32 & 0.0000 \\ 
  party & 6 & 28294.31 & 6 & 923.02 & 0.0000 \\ 
  narrative:party & 6 & 923.02 & 0 & 0.00 & 0.0000 \\ 
   \hline
\end{tabular}
\end{table}

\begin{figure}[htbp]
\centerline{\includegraphics[width=0.7\linewidth]{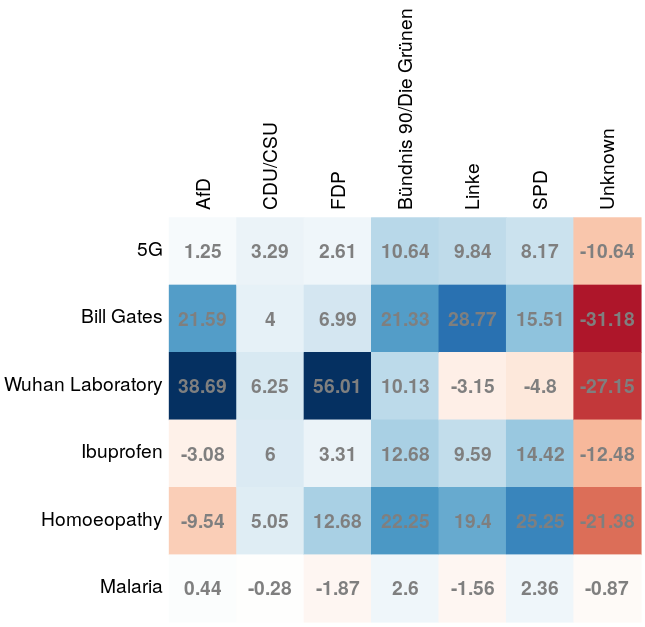}}
\caption{Distribution of residuals of Chi-Square goodness of fit tests}
\label{fig:normalized}
\end{figure}

\noindent
To further estimate the relative effect of political orientation on the contribution to conspiracy narratives on Twitter, we applied six Chi-Square goodness of fit tests on the control group and each of the other six conspiracy narratives. For all of the six tests the p-values were significantly less than 0.05, which suggests that the distributions of the contribution to the six different conspiracy narratives are statistically different compared to the control group. Figure \ref{fig:normalized} shows the distribution of the tests' residuals. The last column of Figure \ref{fig:normalized} shows that the Twitter users without a certain political orientation contributed relatively less to conspiracy narratives in comparison to the control group. In other words, compared to the control group, users with certain political orientations contributed more to the circulation of conspiracy narratives.

\subsection{Research Question 3}

Automated accounts, or users who post programmatically, make up a significant amount of between 9\% and 15\% of Twitter users worldwide \citep{davis2016botornot}. Multiple studies hold automated accounts responsible for political manipulation and undue influence on the political agenda \citep{shao2017spread,ferrara2016rise}. However, more recent studies shed light on these previous results and showed that the influence of automated accounts is overestimated. Ferrara finds that automated accounts comprise less than 10\% of users who post generally about COVID-19 \citep{Ferrara_2020ss}.  

There are multiple methods to automatically detect automated accounts on OSNs \citep{alothali2018detecting}. For this study, we used the method developed by Davis et al. \citep{davis2016botornot}. They applied random forest classification trees on more than a thousand public meta-data available using the Twitter API and on other human engineered features. Table \ref{tab:bot} displays the percentage of automated accounts (users with Complete Automation Probability higher than 0.5) and verified users who contribute to conspiracy narratives. 

\begin{table}[!ht]
\caption{Ratio of tweets posted by automated and verified users}\label{tab:bot}
\begin{tabularx}{\linewidth}{|V{3cm}|X|X|X|X|X|X|X|}
\hline
case             & share of tweets posted by verified users & share of tweets posted by automated accounts & Ratio of automated accounts to verified users \\\hline
5G               & 3.141\%                                  & 1.578\%                        & 0.5                             \\\hline
Bill Gates       & 1.85\%                                   & 1.358\%                        & 0.73                            \\\hline
Wuhan laboratory & 9.065\%                                  & 1.3\%                          & 0.14                            \\\hline
Ibuprofen        & 3.349\%                                  & 1.386\%                        & 0.41                            \\\hline
Homoeopathy      & 1.039\%                                  & 0.921\%                        & 0.89                            \\\hline
Malaria          & 4.626\%                                  & 1.343\%                        & 0.29                            \\\hline
Control group    & 4.644\%                                  & 0.89\%                         & 0.19                            \\\hline
\end{tabularx}
\end{table}

Based on this analysis, 1.31\% of COVID-19 conspiracy narrative tweets are suspected to be posted by automated accounts. This number is significantly lower than many other studies on bot activities on Twitter. We speculate that this occurs due to three reasons. First, the importance of the topic might have captured a lot of public attention, so that significantly more users discuss COVID-19-related topics compared to usual Twitter discussions. Second, many service providers, including Twitter, have started to combat COVID-19 misinformation because of widespread warnings. Finally, we have concentrated on German tweets while the past estimates apply to tweets in English.


\section{Discussions and limitations}
In this study we analyzed more than 9.5M German language tweets and showed that the volume of tweets that discuss one of the six considered conspiracy narratives represents about 0.6\% of all COVID-19 tweets. This translates to more than 36,000 unique German speaking Twitter users. Imhoff and Lamberty found that \quotes{believing that COVID-19 was a hoax was a strong negative prediction of containment-related behaviors like hand washing and keeping physical distance}. To provide the public with accurate information about the importance of such measures, social media intelligence can help elevate potential pitfalls of the Twitter information ecosystem. 

Using more than 38,000 tweets and 36,000 unique Twitter users, we formed the contingency table of political orientation and of contribution to COVID-19 conspiracy narratives (Table \ref{tab:orientation2}). We then applied a saturated Poisson log-linear regression and showed that we cannot statistically reject independence among the underlying variables. This implies partisans have a higher motivation for taking part in COVID-19-related conspiracy discussions. This shows that politically polarized citizens increase the spread of health misinformation on Twitter.

Finally, we employed an automated accounts detection tool and showed that on average about 1.31\% of the users who discuss COVID-19 conspiracy narratives are potentially automated accounts or bots. This number is much lower than estimations on general bot activity on Twitter, which is assumed to be up to 15\% \citep{davis2016botornot,varol2017online}.    

This study holds new insights as well as some limitations: 

\begin{itemize}
\item Our results shed light on the problem of misinformation on Twitter in times of crises for a certain cultural and language context: Germany. We showed that the political orientation of politically polarized users translates to higher circulation of health-related conspiracy narratives on Twitter. Further research could compare the results of this study with other countries and language realms on Twitter. 

\item We also offer indications between political or ideological partisanship and engagement in the dissemination of misinformation on Twitter. In this study we examined if political partisanship motivates individuals to take part in conspiracy discussions. In other words, we did not distinguish between tweets promoting the conspiracy narratives and those rejecting them. One could extend the analysis and study the effect of partisanship on promoting conspiracy theories.  Further research will also need to combine quantitative data analysis and qualitative content analysis to better understand the underlying motivations for engaging in conspiracy communication on OSNs.
  
\item Finally, we offer a more nuanced view on the role of automated tweets regarding a highly emotionally-charged topic. There are numerous studies showing contradictory estimates of bot activity on OSNs. We found only about 1.31\% of users who spread COVID-19 conspiracy tweets are potentially bots. This number is much lower than many of those put forward by other researchers. Further research could investigate this result in order to understand the reasons why this estimation is lower than other case studies.  
\end{itemize}

\bibliographystyle{unsrtnat}
\bibliography{references}  

\end{document}